\documentclass[conference]{IEEEtran}
\usepackage{booktabs}
\IEEEoverridecommandlockouts
\usepackage{cite}
\usepackage{amsmath,amssymb,amsfonts}
\usepackage{algorithmic}
\usepackage{graphicx}
\usepackage{textcomp}
\usepackage{xcolor}
\usepackage{cite}
\usepackage{amsmath,amssymb,amsfonts}
\usepackage{algorithmic}
\usepackage{textcomp}
\usepackage{balance}
\usepackage[T1]{fontenc}
 \usepackage{multirow}

\usepackage[ruled, noend, linesnumbered]{algorithm2e}
\usepackage{graphicx}
\SetKwInput{KwInput}{Input}                
\SetKwInput{KwOutput}{Output}
\usepackage{svg}
\usepackage{tabularx}
\usepackage{multicol}
\usepackage{silence}
\usepackage[font=small, skip=0pt]{subcaption}
\usepackage{textcomp}
\usepackage{xcolor}
\usepackage{float}
\usepackage{comment}
\usepackage{romannum}
\usepackage{ragged2e}

\def\BibTeX{{\rm B\kern-.05em{\sc i\kern-.025em b}\kern-.08em
    T\kern-.1667em\lower.7ex\hbox{E}\kern-.125emX}}

\begin{document}

\pagestyle{empty}
\title{Coherent Optical Quantum Computing-Aided Resource Optimization for Transportation Digital Twin Construction}

\author{
\thanks{
This is a preprint version, full paper has been accepted in IEEE CASCON 2025 and will appear on lEEE Xplore.}
\IEEEauthorblockN{Huixiang Zhang}
\IEEEauthorblockA{\textit{Department of Computer Science} \\
\textit{Lakehead University}\\
Thunder Bay, Canada \\
hzhan102@lakeheadu.ca}
\and
\IEEEauthorblockN{Mahzabeen Emu}
\IEEEauthorblockA{
\textit{Quantum Communications and Computing Research Center} \\
\textit{Department of Electrical and Computer Engineering} \\
\textit{Memorial University of Newfoundland}\\
St. John's, NL, Canada \\
memu@mun.ca}

}

\maketitle

\begin{abstract}
Constructing realistic digital twins for applications such as training autonomous driving models requires the efficient allocation of real-world data, yet data sovereignty regulations present a major challenge. To address this, we tackle the optimization problem faced by metaverse service providers (MSPs) responsible for allocating geographically constrained data resources. We propose a two-stage stochastic integer programming (SIP) model that incorporates reservation and on-demand planning, enabling MSPs to efficiently subscribe and allocate data from specific regions to clients for training their models on local road conditions. The SIP model is transformed into a quadratic unconstrained binary optimization (QUBO) formulation and implemented for the first time at a practical scale on a 550-qubit coherent Ising machine (CIM), representing an exploratory step toward future quantum computing paradigms. Our approach introduces an MSP-centric framework for compliant data collection under sovereignty constraints, a hybrid cost model combining deterministic fees with probabilistic penalties, and a practical implementation on quantum hardware. Experimental results demonstrate that CIM-based optimization finds high-quality solutions with millisecond-scale ($10^3$ second) computation times, significantly outperforming quantum-inspired solvers like PyQUBO. Although classical solvers such as Gurobi can achieve marginally better solution quality, CIM is orders of magnitude faster, establishing a practical paradigm for quantum-enhanced resource management.
\end{abstract}


\section{Introduction}

The metaverse refers to the intersection of physical and virtual space to create a hybrid reality supported by ubiquitous communication and computing infrastructure.
\subsection{Background}
The rapid development of the Metaverse is driven by advancements in key technologies. The proliferation of Internet of Things (IoT) devices, alongside 5G networks and edge computing, has created a data-intensive environment supporting real-time, responsive interactions. This technological progress is fueling significant market growth, with revenue projected to reach \$936.6 billion by 2030, accompanied by an almost twenty-fold increase in data usage \cite{zotero-120, ng2021}. This surge in data volume makes efficient resource allocation a critical challenge for Metaverse data providers, setting the stage for our work.
\begin{figure*}[t]
    \centering
    \includegraphics[width=.9\textwidth]{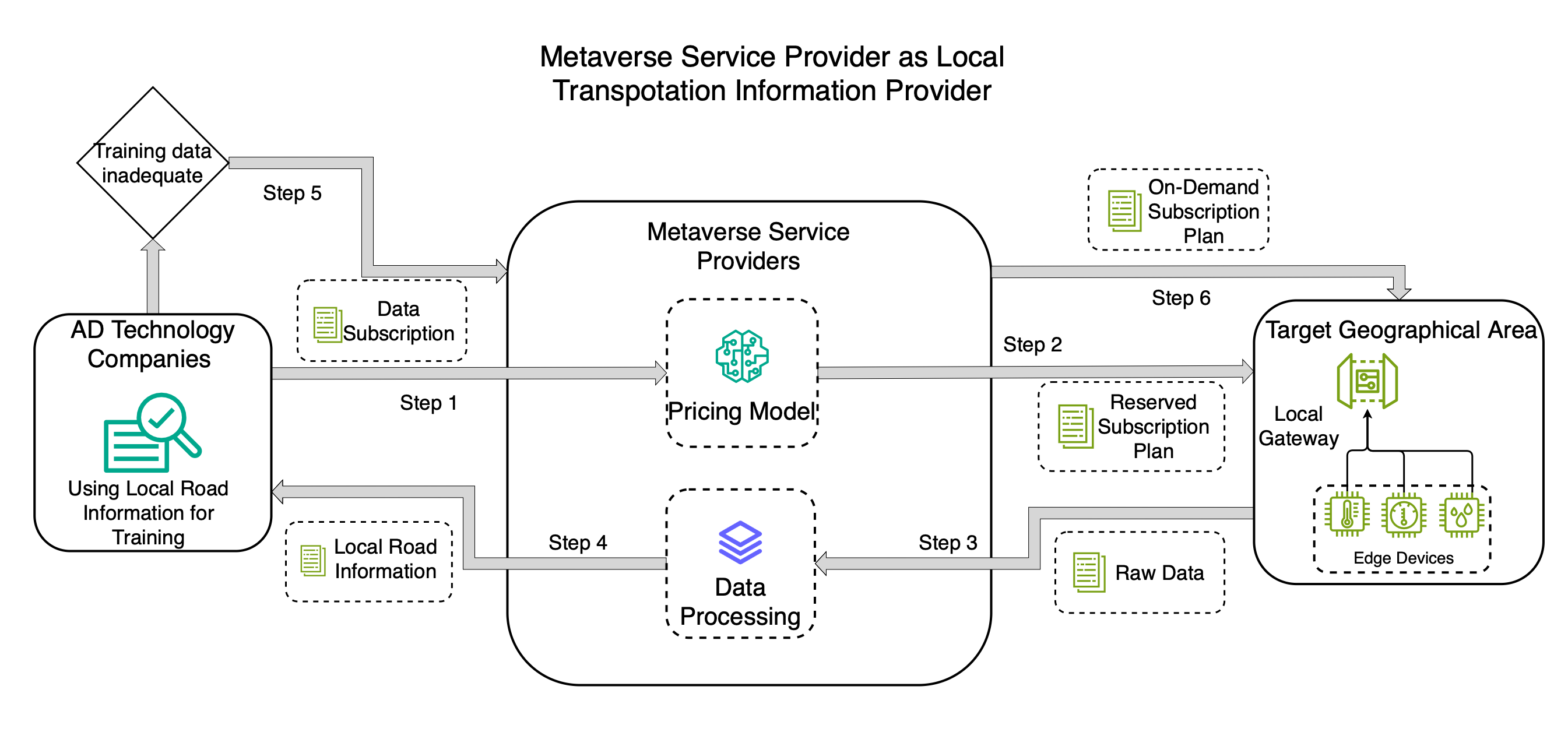}
    \vspace{-.2CM}\caption{The workflow of Metaverse Service Provider with two-phase training data.}
    \label{fig:your_label} 
\end{figure*}

\subsection{Research Contributions}
In this paper, we propose a digital twin system that provides real-time traffic condition mapping for specific geographic areas. Due to regulatory restrictions (e.g. concerns about road data security), foreign autonomous driving technology companies often face challenges in obtaining local training permits, limiting their ability to adapt models to regional traffic patterns. Our quantum-based digital twin solution aims to dramatically lower location-specific training costs and enhance real-time adaptability.

Inspired by previous literature \cite{ng2024, emu2024}, we confine the Metaverse Service Provider (MSP) to an entity that provides virtual training data to autonomous driving companies, while edge devices act as sensors collecting real-time data from target geographical regions and design an MSP-centric workflow specifically used to obtain transportation information in the target geographical area (TGA). As illustrated in Fig. 1, the workflow begins when autonomous driving (AD) technology companies submit service requests to the MSP, which employs a pricing model to minimize training costs. Subsequently, the MSP sends a Reserved Subscription Plan to the TGA, enabling the reserving of edge devices for the collection of raw data. These edge devices transmit real-time raw data to the MSP (Step 3), which processes them into structured local transportation information for delivery to AD companies (Step 4). If the training data are inadequate, AD companies issue supplemental requests (Step 5), and ask the pricing model to generate an On-Demand Subscription Plan (Step 6). This plan follows the same data loop (Steps 3–4), closing the iterative feedback cycle.

To minimize costs, we propose a customized two-stage stochastic integer programming (SIP) model inspired by \cite{emu2024}. The model is transformed into a quadratic unconstrained binary optimization (QUBO) problem, which is solvable via both physical quantum machines such as coherent Ising machines (CIM) and quantum-inspired methods in scientific computing libraries like PyQUBO. Our contributions are summarized as follows:

\begin{itemize}
\item \textbf{Workflow Architecture}: Design an MSP-centric framework that integrates reserved and on-demand subscription plans, enabling dynamic data collection from edge devices in regulated TGAs while maintaining compliance with data sovereignty constraints.
\item \textbf{Parameter Simplification}: Optimize SIP parameters to prevent excessive growth in matrix dimensions during the QUBO solve process. 
\item \textbf{Quantum Implementation}: Demonstrate the potential of quantum hardware in handling exponentially growing problem dimensions through practical solving, paving the way for quantum computing adoption in the digital twin within Metaverse domain.
\end{itemize}

\subsection{Paper Structure}
The remainder of this paper is organized as follows. Section II reviews related work. Section III formulates the problem using a two-stage stochastic integer programming (SIP) framework. Section IV presents the transformation of the SIP formulation into a quadratic unconstrained binary optimization (QUBO) problem. Section V conducts simulations and analyzes the results. Finally, Section VI concludes the paper.

\section{Related Work}
This review covers three key areas: resource allocation in the metaverse, digital twin systems for transportation, and the application of quantum optimization to these problems.

A primary challenge in the metaverse is the efficient allocation of resources to bridge the physical and virtual worlds. Previous work has focused on establishing foundational frameworks for IoT coordination and improving resource discovery efficiency \cite{ning, nunes2018}. Optimization efforts have employed classical methods such as mixed integer linear programming (MILP) to minimize latency in edge computing \cite{mao2017} and user-centric schemes to improve quality of experience (QoE) \cite{du2022}. However, these approaches often assume deterministic conditions or use static resource partitioning, which struggles with the real-time demand fluctuations inherent in dynamic environments \cite{moorthy2020, mao2017}. More recently, quantum computing has emerged to address these limitations. For example, a hybrid classical-quantum framework using Warm Start Quantum Annealing (WSQA) demonstrated a 40\% cost reduction under stochastic demands compared to traditional MILP solvers, highlighting the potential of quantum approaches \cite{emu2023}.

In parallel, digital twin systems have become critical for advancing autonomous driving (AD) by overcoming the biases and edge case limitations of public datasets \cite{caesar2020}. Research in this area aims to generate cost-effective and scalable training data that accurately replicate real-world environments. Recent frameworks have introduced realistic multi-agent interactions to mitigate simulation overfitting \cite{gulino2023} and used OpenStreetMap (OSM) data to dynamically generate closed-loop traffic simulations for any city, offering unprecedented flexibility without prebuilt 3D assets \cite{yang2024}. 

The most promising quantum approaches for such optimization tasks involve solving Quadratic Unconstrained Binary Optimization (QUBO) problems on specialized hardware like quantum annealers or Coherent Ising Machines (CIMs). These methods leverage quantum dynamics to explore many potential solutions in parallel, and their hybrid integration with classical systems has shown superior performance under the uncertain conditions typical of metaverse applications \cite{emu2023}.

Building upon the technical limitations identified in Section II, we highlight two unresolved issues in Metaverse resource allocation: (1) Absence of semantic-aware identification of client requirements for reserving and allocating real-time data streams tailored to specific geographical environments or scenario types. Existing frameworks lack mechanisms to deliver spatially targeted data services. (2) Deficiency in adaptive QUBO modeling capabilities to address real-time demand fluctuations. Current static formulations fail to automatically adjust to dynamic optimization landscapes, leading to suboptimal resource utilization under time-varying conditions. 

These critical gaps directly motivate the architectural design of our system model. Leveraging next-generation quantum computing paradigms, we propose a framework that enables user-customizable digital twins can reflect real-time traffic conditions in specified geographical areas. The system further incorporates algorithmic optimization of edge device subscription plans to reduce operational costs for MSPs in acquiring high-fidelity environmental data.

\begin{table}[h]
    \centering
    \caption{Parameters and Decision Variables for Resource Subscription Model}
    \label{tab:parameters}
    \renewcommand{\arraystretch}{1.2}
    \begin{tabular}{|p{1.8cm}|p{6cm}|}
        \hline
        \multicolumn{2}{|c|}{\textbf{System Model Parameters}} \\
        \hline
        \textbf{Notation} & \textbf{Description} \\
        \hline
        $\mathcal{W}$ & Set of Metaverse Service Providers\\
        $\Omega$ & Set of possible scenarios \\
        $\mathcal{E}$ & Set of edge devices \\
        $X$ & Maximum reserved bundles per subscription \\
        $C_e^{(r, \text{memb})}$ & Membership cost per edge \\
        $C_e^{(r, \text{trans})}$ & Reserved data bundle transmission cost per edge\\
        $C_e^{(o, \text{trans})}$ & On-demand data bundle transmission cost per edge\\
        \hline
        \multicolumn{2}{|c|}{\textbf{Scenario Parameters}} \\
        \hline
        \textbf{Notation} & \textbf{Description} \\
        \hline
        $\lambda_s$ & Scenario index \\
        $P(\lambda_s)$ & Probability of scenario $\lambda_s$ \\
        $S_{w,e}(\lambda_s)$ & Similarity score in scenario $\lambda_s$ \\
        $\bar{F}_w(\lambda_s)$ & Actual demand of MSP $w$ in $\lambda_s$ \\
        \hline
        \multicolumn{2}{|c|}{\textbf{Decision Variables}} \\
        \hline
        \textbf{Notation} & \textbf{Implementation Details} \\
        \hline
        $m_{w,e}^{(r)}$ & Binary subscription indicator \\
        $\tilde{m}_{w,e}^{(r)}$ & Reserved bundles purchased\\
        $m_{w,e}^{(o)}(\lambda_s)$ & On-demand bundles in scenario $\lambda_s$ \\
        \hline
    \end{tabular}
\end{table}

\section{System Model}
In this section, we formulate a two-phase cost minimization problem addressing both reservation and on-demand subscription scenarios. The objective function explicitly handles uncertainty. In formulating the SIP model, we proactively consider the exponential variable growth challenge for further QUBO conversion by implementing a low-coupling parameter design. This strategic parameterization prevents the explosion of the variable beyond the magnitude of the $10^{3}$ order in subsequent quantum optimization stages.

The objective function consists of costs from two stages: a deterministic reservation phase and a stochastic on-demand phase. The first part of the objective, $\sum_{w \in \mathcal{W}} \sum_{e \in \mathcal{E}} \left( m_{w,e}^{(r)} C_e^{(r, \text{memb})} + \tilde{m}_{w,e}^{(r)} C_e^{(r, \text{trans})} \right)$, represents the first-stage reservation cost. In this stage, MSPs purchase memberships for specific edge devices and pay for a reserved number of data bundle transmissions. Notably, the data bundle encompasses the entire process of data collection, semantic extraction, and data transformation, with the core semantic extraction carried out by the edge device providers.

\[
\min_{m_{w,e}^{(r)}, \tilde{m}_{w,e}^{(r)}, m_{w,e}^{(o)}(\lambda_s)}:
\]

\[
\sum_{w \in \mathcal{W}} \sum_{e \in \mathcal{E}} \left( m_{w,e}^{(r)} C_e^{(r, \text{memb})} + \tilde{m}_{w,e}^{(r)} C_e^{(r, \text{trans})} \right)
+ \mathbb{E} \left[ Q \left( m_{w,e}^{(o)}(\lambda_s) \right) \right], 
\]
where 
\[
Q \left( m_{w,e}^{(o)}(\lambda_s) \right) = \sum_{\lambda_s \in \Omega} P(\lambda_s) \times \sum_{w \in \mathcal{W}} \sum_{e \in \mathcal{E}} m_{w,e}^{(o)}(\lambda_s) C_e^{(o, \text{trans})}
\]

The second stage of the model addresses the uncertainty inherent in the demand for real-world data. This stochastic nature of the problem is captured through a set of discrete scenarios $\Omega$. Each scenario $\lambda_s \in \Omega$ represents a distinct and possible future state of the transportation environment. For example, a scenario could be 'heavy morning rush-hour traffic under clear weather' versus another scenario of 'light weekend traffic during a snowstorm.' The probability of each scenario, $P(\lambda_s)$, is estimated from historical traffic and meteorological data, with $\sum_s P(\lambda_s) = 1$. The parameters dependent on these scenarios, such as the total data demand $\bar{F}_w(\lambda_s)$ and the data similarity score $S_{w,e}(\lambda_s)$ (e.g., how relevant a camera's view is during a specific event), are consequently defined for each unique state.

This leads to the second part of the objective function, which calculates the expected on-demand cost across all scenarios: $\sum_{\lambda_s \in \Omega} P(\lambda_s) \times \sum_{w \in \mathcal{W}} \sum_{e \in \mathcal{E}} m_{w,e}^{(o)}(\lambda_s) C_e^{(o, \text{trans})}$. Here, $m_{w,e}^{(o)}(\lambda_s)$ is the number of on-demand data bundles purchased in a specific scenario $\lambda_s$, and this cost is weighted by the probability of the scenario $P(\lambda_s)$, ensuring that the total objective minimizes the expected cost under uncertainty.

The SIP model is subject to the following constraints.
\[
\mathcal{C}1: \tilde{m}_{w,e}^{(r)} \leq m_{w,e}^{(r)} X, \quad \forall w \in \mathcal{W}, \forall e \in \mathcal{E}
\]
The constraint $C1$ that subscription fees must be paid before the purchase of any data bundle, $m_{w,e}^{(r)} = 1$. Additionally, even after subscribing, the MSP can purchase at most $X$ reserved data bundles from that edge device.

\[
\mathcal{C}2: \sum_{e \in \mathcal{E}} \tilde{m}_{w, e}^{(r)} S_{w,e}(\lambda_s) + \sum_{e \in \mathcal{E}} m_{w,e}^{(o)}(\lambda_s) \geq \bar{F}_w(\lambda_s),\]
\[\quad \forall w \in \mathcal{W}, \forall \lambda_s \in \Omega
\]

The constraint $C2$ ensures that, in each scenario $\lambda_s$, the total amount of data (from both reserved and on-demand purchases) meets or exceeds the MSP's demand if the MSP deems it acceptable $\bar{F}_w(\lambda_s)$. The term $\tilde{m}_{w, e}^{(r)} S_{w,e}(\lambda_s)$ represents the effective number of data bundles (e.g., images), adjusted by the similarity score $S_{w, e}(\lambda_s)$, which reflects how well the data from the edge device $e$ matches the MSP requirements in scenario $\lambda_s$.

\[
\mathcal{C}3: m_{w,e}^{(r)} \in \{0,1\}, \quad \forall w \in \mathcal{W}, \forall e \in \mathcal{E}
\]

Constraint $C3$ enforces binary subscription decisions, where $m_{w,e}^{(r)} = 1$ indicates that the edge device $e$ is subscribed by the MSP $w$ in the reservation phase.

\[
\mathcal{C}4: \tilde{m}_{w,e}^{(r)} \in \mathbb{Z}^+, \quad \forall w \in \mathcal{W}, \forall e \in \mathcal{E},
\]

Constraint $C4$  ensures that the number of reserved data bundles purchased is a non-negative integer.
\[
\mathcal{C}5: m_{w,e}^{(o)}(\lambda_s) \in \mathbb{Z}^+, \quad \forall w \in \mathcal{W}, \forall e \in \mathcal{E}, \forall \lambda_s \in \Omega. 
\]

Constraint $C5$ ensures that the number of on-demand data bundles purchased in any scenario is a non-negative integer.

\section{QUBO Formulation}

The QUBO formulation is a universal representation fit for solving combinatorial optimization problems on both physical quantum annealing devices and quantum-inspired solvers. In this section, we present the binary encoding scheme and penalty-based constraint embedding techniques that transform the previously developed two-stage SIP model into an equivalent QUBO formulation.

\subsection{Quantum Annealing Fundamentals and Encoding Necessity}
QUBO serves as the canonical formulation for combinatorial optimization in quantum annealing devices. The major technical challenge is QUBO only works with binary decision variables, whereas real-time decision making such as digital twin construction in Metaverse and its purchase schemes involve continuous variables. 
To bridge the gap between QUBO's binary variables ($b \in \{0,1\}$) and continuous decision variables in resource allocation (e.g., $\tilde{m}_{w,e}^{(r)} \in \mathbb{Z}^+$), we propose a hierarchical encoding mechanism. The binary encoding mechanism translates continuous variables into binary representations compatible with quantum computing processes, implemented through $K$-bit binary expansion ($K = \lceil \log_2 X \rceil$) while maintaining 8-bit hardware compatibility. The core encoding relationships are defined as follows:

\begin{equation}
    \tilde{m}_{w,e}^{(r)} = \sum_{k=1}^K 2^{k-1} b_{w,e,k}^{(r)}, \quad
    m_{w,e}^{(o)}(\lambda_s) = \sum_{l=1}^L 2^{l-1} b_{w,e,l}^{(o)}(\lambda_s)
\end{equation}
where \( b_{w,e,k}^{(r)}, b_{w,e,l}^{(o)} \in \{0,1\} \) are binary decision variables. Adaptive precision tuning and constraint-preserving penalty coefficients achieve a three-way balance between solution quality (quantization error \( <4\% \) when \( K=5 \)), computational efficiency, and objective comparability with classical models. This scheme reduces the problem dimensionality from $O(|\mathcal{W}| \times |\mathcal{E}| \times |\lambda|) \ \text{(continuous variables)}$ to $O(|\mathcal{W}| \times |\mathcal{E}| \times (K + |\lambda| \times K)) \ \text{(binary variables)}$, directly mapped to the quantum processor's physical implementation. The reduction stems from encoding integer variables via compact binary representations \( \sum_{k=0}^{K-1} 2^k b_k \), while adaptive constraints and penalty coefficients maintain equivalence with classical continuous optimization frameworks under bounded quantization error.

\subsection{QUBO Model Construction}
The complete QUBO formulation combines objective costs with quadratic penalty terms for constraint enforcement:

\begin{equation}
    \begin{aligned}
        H = {} 
        &\underbrace{\sum_{w,e} \left[ C_e^{(r,\text{memb})} m_{w,e}^{(r)} + C_e^{(r,\text{trans})} \tilde{m}_{w,e}^{(r)} \right]}_{\text{Membership costs}} \\
        &+ \underbrace{\mathbb{E}_\lambda \left[ \sum_{w,e} C_e^{(o,\text{trans})} m_{w,e}^{(o)}(\lambda) \right]}_{\text{On-demand costs}} 
        + \underbrace{\alpha P_{C_1} + \beta P_{C_2}}_{\text{Constraints}}
    \end{aligned}    
\end{equation}

The constraint terms implement the problem's physical requirements through quadratic penalties:
\begin{align}
    P_{C_1} &= \sum_{e \in E} \tilde{m}_{w,e}^{(r)} \leq X \sum_{e \in E} m_{w,e}^{(r)}, \quad \forall w \in W \\
    P_{C_2} &= \sum_{w,\lambda_s} \left( \bar{F}_w(\lambda_s) - \sum_e \left[ S_{w,e}(\lambda_s)\tilde{m}_{w,e}^{(r)} + m_{w,e}^{(o)}(\lambda_s) \right] \right)^2
\end{align}

The penalty coefficients $\alpha$ and $\beta$ are critical to enforcing constraints in a QUBO model. Their values must be large enough to make any violation of a constraint energetically unfavorable, ensuring that the low-energy solutions found by the solver are valid. Following common practice in QUBO formulation, we set these coefficients to be significantly higher than any potential cost in the objective function to ensure constraint dominance.

Through empirical tuning, we selected $\alpha=10000$ and $\beta=100$. This hierarchical setting, where the penalty for the fundamental subscription logic (C1) is much greater than for the service demand (C2), ensures that all solutions returned from the CIM were feasible while also preserving a reasonable structure for the quantum energy landscape. Although these values proved to be effective, a detailed sensitivity analysis of these parameters is a potential avenue for future work.

The final model contains $O(|\mathcal{W}||\mathcal{E}|(K + |\Omega|L))$ binary variables, directly corresponding to the $C_e^{(r, \text{trans})}$ and $C_e^{(o, \text{trans})}$ arrays in the implementation.

\section{Performance Analysis}
Throughout this section, we describe the experiment environment and parameter configurations considered in our experiments. We then demonstrate our model's capability to optimize data demand processing as an intermediate layer between edge devices and end-users. Subsequent experiments employing a Coherent Ising Machine (CIM), Gurobi, and PyQUBO reveal that solving identical QUBO models on a CIM reveals a significant performance trade-off. While CIM provides solutions orders of magnitude faster, the classical Gurobi solver achieves marginally better solution quality, as we will demonstrate.

We initially formulated the problem as an SIP model using the Gurobi optimizer. After converting it to a QUBO formulation, we performed simulated computations using both the solve\_qubo method from gurobi\_optimods (classical) and the simulated annealing approach provided by PyQUBO (quantum-inspired). 

For physical experiments, we formulated the QUBO model in Python and utilized the Kaiwu SDK to interface directly with the CPQC-550, a 550-qubit coherent photonic quantum computer from QBoson. The SDK handled the mapping of our QUBO matrix onto the quantum processing unit and initiated the physical annealing process, with the results returned for classical analysis. This represents a direct execution on quantum hardware, not a classical simulation of a quantum computer.

Three distinct problem scales (small, medium, and large) were simulated with systematic parameter configurations as summarized in Table II. The potential scenario range was fixed between 2 and 5, while the quantity of edge devices ranged from 5 to 20. The total demand for data bundles followed a Poisson distribution. The cost ratios of subscription fee, reserved transmission and on-demand purchase were set at 10: 1: 5 based on empirical market data \cite{zhang2023}.

\begin{table}[ht]
\caption{Parameter Configurations Across Problem Scales}
\centering
\begin{tabular}{lccc}
\toprule
\textbf{Parameter} & \textbf{Small (S)} & \textbf{Medium (M)} & \textbf{Large (L)} \\
\midrule
Number of MSPs & 1 & 2 & 5 \\
Edge Devices & 5 & 10 & 25 \\
Service Scenarios & 2 & 3 & 5 \\
Data Bundles Demand & 2,000 & 6,000 & 15,000 \\
\bottomrule
\end{tabular}
\label{tab:params}
\end{table}

\subsection{Compararison of CIM Against Gurobi and PyQUBO}
To evaluate the potential of optical quantum computing in communication systems, we implemented our Metaverse content resource allocation model using the CIM, Gurobi, and PyQUBO. Fig. 2 compares the solution quality of these three approaches on different problem scales. The results highlight a trade-off between computational time and solution optimality. As shown in Fig. 2, the classical Gurobi optimizer consistently found the lowest objective values, indicating the highest quality solutions across all problem scales. However, the Coherent Ising Machine (CIM) produced highly competitive solutions, achieving objective values that were only marginally higher than those of Gurobi's, while both significantly outperformed the PyQUBO simulated annealer.

As shown in Fig. 3 and Table II, the CIM maintains a millisecond-scale computation time across all problem sizes, with execution times of 0.001718s (S-scale), 0.001687s (M-scale), and 0.005235s (L-scale). The slight time reduction from small to medium scale is likely attributable to system overhead and measurement fluctuations in such short execution windows. Critically, the observed absence of exponential runtime growth relative to problem size fundamentally distinguishes the CIM from classical and quantum-inspired solvers, establishing its capability for real-time decision-making in dynamic metaverse environments.

\begin{table}[htbp]
    \centering
    \caption{Execution Time Comparison Across Problem Scales}
    \label{tab:runtime}
    \begin{tabular}{lccc}
        \toprule
        \textbf{Method} & \textbf{S-scale (s)} & \textbf{M-scale (s)} & \textbf{L-scale (s)} \\
        \midrule
        CIM             & 0.001718           & 0.001687           & 0.005235           \\
        Gurobi          & 0.0311             & 0.0674             & 1.5235             \\
        PyQUBO          & 1.42               & 13.78              & 310.765579         \\
        \bottomrule
    \end{tabular}
\end{table}

As shown in Table III, the execution time of CIM under different problem sizes %
remains at the millisecond level, while the execution time of Gurobi shows a linear %
increase with the increase of problem size, and the execution time of PyQUBO increases exponentially.

\begin{figure}[htp]
    \centering
    \includegraphics[width=\linewidth]{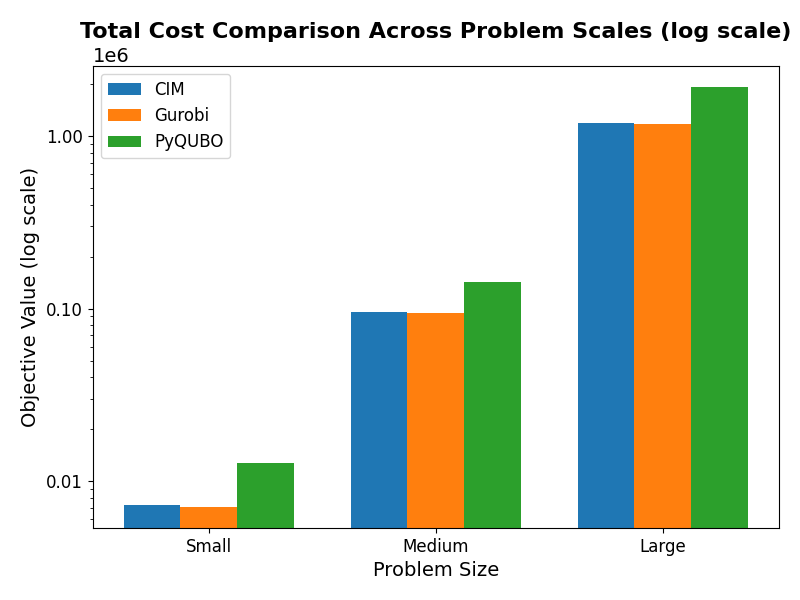}
    \caption{Comparison of logarithmic total costs obtained via CIM (quantum hardware), Gurobi, and PyQUBO under small, medium, and large problem configurations defined in Table II.}
    \label{fig:methods_comparison}
\end{figure}

\begin{figure}[htp]
    \centering
    \includegraphics[width=\linewidth]{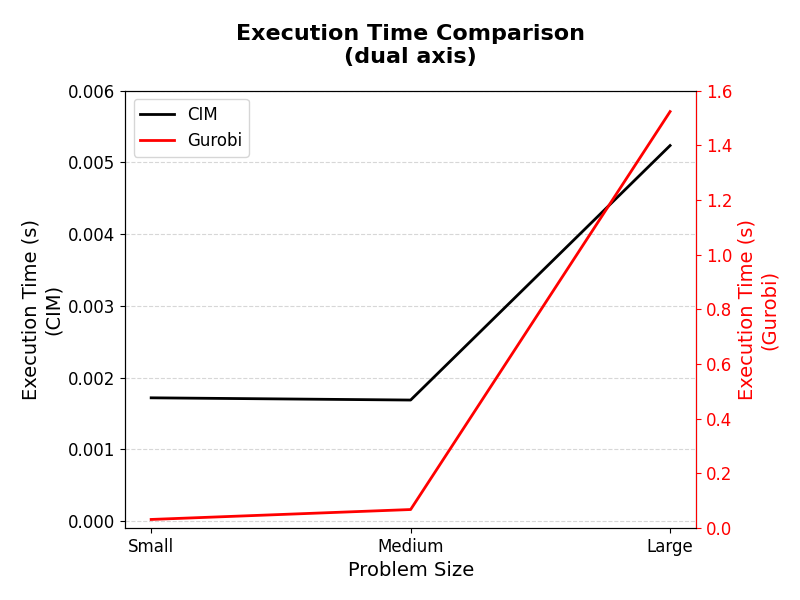}
    \caption{Comparison of computation time between coherent Ising machine(left axis) and Gurobi(right axis) for solving the target equation across small, medium, and large problem scale.}
    \label{fig:runtime_compare}
\end{figure}

\subsection{Performance Analysis of Quantum Method}
The numerical experiments in Fig. 2 demonstrate that the proposed quantum optimization model effectively addresses dynamic resource allocation in metaverse data markets. By introducing binary encoding strategies that map multidimensional decision variables ($m_{w,e}^{(r)}$, $\tilde{m}_{w,e}^{(r)}$, $m_{w,e}^{(o)}(\lambda_s)$) to compact QUBO representations, reduced problem dimensionality.

Specifically, the bit decomposition technique implemented in the code, %
which restricts $K=min(ceil(log_2(X+1)),5)$, significantly reduces the size of the QUBO matrix. %
This encoding mechanism inherently aligns with the superposition principle of quantum computing, %
preserving exponential parallelism in parameter space exploration.

These results conclusively demonstrate the practical viability of our quantum-enhanced framework. While classical solvers exhibit comparable optimality gaps (<5\%) in small-scale scenarios, they suffer from prohibitive exponential time complexity growth as problem dimensions increase. In contrast, our approach maintains polynomial time complexity $(O(n^2))$ at different scales through hardware-aligned QUBO encoding, achieving high constraint satisfaction rates at metro-scale deployments — a critical advancement for real-time Metaverse resource orchestration. Current limitations primarily stem from hardware constraints, including restricted qubit connectivity in superconducting annealers and optical CIMs' sensitivity to QUBO-to-Ising mapping precision for Metaverse-specific constraints, alongside emerging challenges in handling heterogeneous data availability with coverage gaps.


\section{Conclusion}
This paper addresses the complex optimization problem of dynamic resource allocation in Metaverse data markets by %
proposing a quantum-inspired mixed-integer linear programming model. By integrating edge device subscription decisions %
with stochastic demand satisfaction mechanisms, we construct a QUBO model incorporating quadratic %
penalty terms, and pioneer its implementation on a 550-qubit CIM for large-scale practical problem solving. %
Numerical experiments demonstrate that CIM exhibits significant advantages in solution quality and computational efficiency compared to %
traditional Gurobi optimizer and PyQUBO simulated annealing approaches. Maintaining millisecond-level response times despite 300\% data volume %
growth, CIM's near-linear time complexity effectively overcomes the ``curse of dimensionality'' that limiting classical methods. %
Our K-bit binary encoding strategy maps multidimensional decision variables to compact QUBO representations, reducing %
problem dimensionality to $O(|\mathcal{W}| \times |\mathcal{E}| \times (K + |\lambda| \times K))$ while maintaining solution accuracy. 
Although current quantum hardware has limitations on the number of qubits, this work validates optical quantum computing's potential in communication resource allocation. %
Future research may explore dynamic penalty coefficient adjustment mechanisms and distributed quantum-classical %
hybrid architectures to address more complex multi-objective optimization scenarios.

\bibliographystyle{IEEEtran}
\bibliography{References} 
\balance

\end{document}